# Impact of the Ligand Shell on Structural Changes and Decomposition of All-Inorganic Mixed-Halide Perovskite ($CsPbX_3$) Nanocrystals under X-Ray Illumination

*Jan Wahl*[$,‡], *Philipp Haizmann*[$,‡], *Christopher Kirsch*[$], *Rene Frecot*[$], *Nastasia Mukharamova*[#], , *Dameli Assalauova*[#], *Young Yong Kim*[#], *Ivan Zaluzhnyy*[+], *Thomas Chassé*[$,ß], *Ivan A. Vartanyants*[#,&], *Heiko Peisert*[$,]*, *Marcus Scheele*[$,ß,]*

[$] Institut für physikalische und theoretische Chemie, Universität Tübingen, Auf der Morgenstelle 18, 72076 Tübingen, Germany

[#] Deutsches Elektronen-Synchrotron DESY, Notkestraße 85, 22607 Hamburg, Germany

[+] Institut für Angewandte Physik, Universität Tübingen, Auf der Morgenstelle 10, 72076 Tübingen, Germany

[ß] Center for Light-Matter Interaction, Sensors & Analytics LISA[+], Universität Tübingen, Auf der Morgenstelle 15, 72076 Tübingen, Germany

[&] National Research Nuclear University MEPhI (Moscow Engineering Physics Institute), Kashirskoe shosse 31, 115409 Moscow, Russia

[‡] These authors contributed equally

*To whom correspondence should be addressed: marcus.scheele@uni-tuebingen.de and heiko.peisert@uni-tuebingen.de

ABSTRACT

We show that the decomposition of caesium lead halide perovskite nanocrystals under continuous X-ray illumination depends on the surface ligand. For oleic acid/oleylamine, we observe a fast decay accompanied by the formation of elemental lead and halogen. Upon surface functionalization with a metal porphyrin derivate, the decay is markedly slower and involves the disproportionation of lead to $Pb^0$ and $Pb^{3+}$. In both cases, the decomposition is preceded by a contraction of the atomic lattice, which appears to initiate the decay. We find that the metal porphyrin derivative induces a strong surface dipole on the nanocrystals, which we hold responsible for the altered and slower decomposition pathway. These results are important for application of lead halide perovskite nanocrystals in X-ray scintillators.

Lead halide perovskites are an important class of materials for use in light harvesting and light emitting devices. [1,2] Another promising application of perovskites are scintillators, where they have already shown good performance in X-ray detection. [3–5] However, the instability of perovskite-based materials is a significant drawback for the application in optoelectronic and scintillating devices. [4,6–11] To address this issue, the photodegradation of methylammonium lead iodide ($MAPbI_3$) thin films has been thoroughly investigated, including the postulation of a decay mechanism. [12–14] Recently, these photodegradation studies have been extended to caesium mixed-halide perovskites ($CsPbX_3$) [15–17] in view of their improved stability against long-term light exposure. [18] However, even the most stable mixed halide perovskites, such as $CsPbBrI_2$, undergo slow photodegradation in intense visible light. [15–17] The aim of this work is to use the tunable surface chemistry of nanocrystals (NCs) to further mitigate this instability with a particular focus on photodegradation with X-ray photons to aid the application in scintillators. [8,13]To this end, we introduce zinc-(5-monocarboxyphenyl-10,15,20-triphenylporphyrin) (mZnTPP) as surface ligand and investigate its effect on the decomposition of the NCs under X-ray illumination. As NCs, we use two model systems, namely $CsPbBrI_2$ and $CsPbBr_2Cl$, based on their light emitting properties hereafter referred to as red and blue perovskites, respectively. If not explicitly stated otherwise, these NCs are surface-functionalized with oleic acid/oleylamine, referred here to as the "native ligand". As previously reported, perovskites experience drastic changes under continuous X-ray exposure. [19,20] One analytical method that utilizes X-ray irradiation is X-ray photoelectron spectroscopy (XPS). Here, a sample is continuously irradiated with X-rays, and electrons, released from the sample due to the external photoelectric effect, are detected based on their kinetic energy. The core-level binding energy of an emitted electron is directly related to the detected kinetic energy. Previous XPS experiments on perovskites reported a shift of the binding energies to higher values with ongoing measurement time. This was mostly attributed to either surface, substrate or charging effects. [7] In this work, we observe a similar shift for the red and blue perovskite NCs. However, we demonstrate by a combination of XPS and wide-angle X-ray scattering (WAXS) that the core-level energy shift originates from a contraction of the atomic lattice. For the red and blue perovskite NCs with native ligands, we find a similar decay mechanism as recently postulated for thin films. [17] While morphology and chemical composition are maintained in the red perovskites, the blue perovskites exhibit a loss of chlorine and undergo recrystallization. Most importantly, we observe a greatly altered and slower decay mechanism after surface functionalization with the porphyrin derivative mZnTPP.

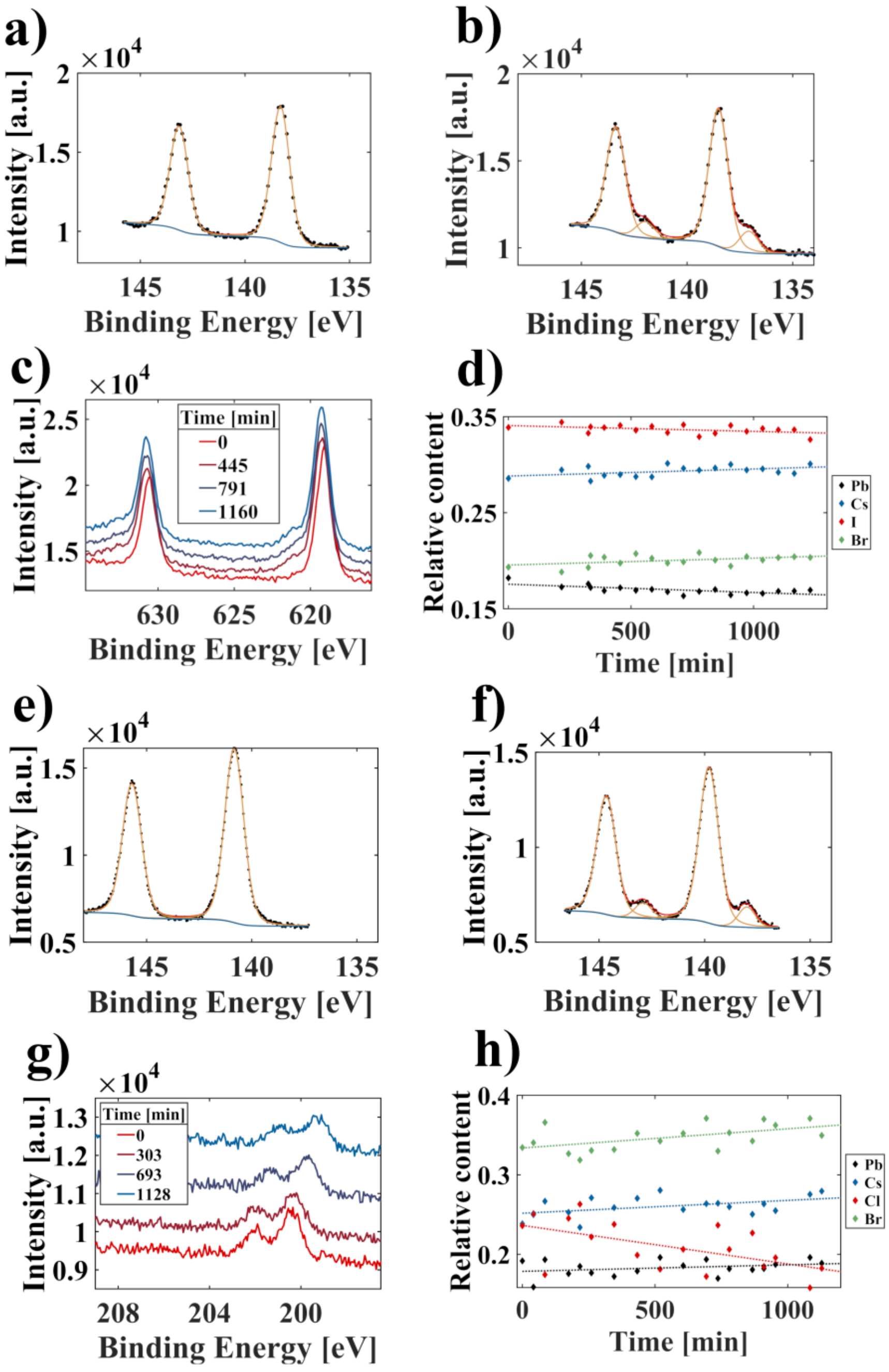

**Figure 1.** Pb 4f core level XPS spectra of $CsPbBrI_2$ a) at the start (0 min) and b) after 1160 min of X-ray exposure. Fit components are shown in orange, the overall fit is depicted in red. All spectra were fitted by applying a Shirley-type background (blue). c) I 3d spectra at different times during the illumination with the evolution of a shoulder at higher binding energy (621 eV). d) Relative stoichiometric content and linear regression (dotted lines) during the exposure for $CsPbBrI_2$. Pb 4f spectra of $CsPbBr_2Cl$ e) at the start (0 min) and f) after 1128 min of exposure. g) Cl 2p spectra (~ 200 eV) exhibit a decrease over time without the formation of a novel peak. h) Relative stoichiometric content for $CsPbBr_2Cl$ during the experiment.

The decomposition of both perovskite systems was analyzed under continuous X-ray illumination for at least 20 h by XPS and ultraviolet photoelectron spectroscopy (UPS). Core level and valence spectra were recorded at regular time intervals. Exemplarily, spectra of Pb 4f doublet peaks recorded at the beginning and end of the illumination are shown in Figure 1a) + b). The Pb 4f main peak at a binding energy of 138.25 eV in Figure 1a) corresponds to $Pb^{2+}$ in the red perovskites. Over the course of this investigation, all samples, independent of stoichiometry and ligand shell, developed an additional Pb species visible in the corresponding Pb 4f-core level spectra of Figure 1a) + b). The signal occurred at lower binding energies (~137 eV), which indicates a lowering of the oxidation state. Therefore, and in agreement with literature, we attribute this peak to the formation of elemental lead ($Pb^0$). [7,12,21] The formation of the $Pb^0$ component is detected after illuminating the sample for few hours. The intensity of the $Pb^0$ peak increases continuously with time, the relative $Pb^0/Pb^{2+}$ content as a function of the illumination time is summarized in Figure 1d) for the red perovskites. Corresponding peak fits are shown for the first and last measurement of the experiment in Figure 1a) + b), respectively. Additionally, we tracked the lead content over the course of the experiment, as shown in Figure S3. After 20 h, we observe the formation of 13.2 % $Pb^0$ for the red perovskites (Figure S3). As a source of electrons for the reduction of $Pb^{2+}$, we consider the halides in the NCs. While the bromide peak in all samples experienced no significant change (see Figure S1), we find strong changes in the core level spectra of the other halides, *i.e.*, $I^-$ for the red perovskites (Figure 1c) + g). We observe the formation of a distinct shoulder at high binding energy in the I 3d region for the red perovskites, suggesting that $I^-$ is indeed oxidized during the reduction of $Pb^{2+}$ (Figure 1c). By tracking the stoichiometric composition, we find the iodide content to stay constant throughout the experiment (Figure 1d), suggesting that the new iodine/iodide species remains in the sample during the decomposition even in ultra-high vacuum. The ratio of the novel iodine (621 eV) to $Pb^0$ (137 eV) species is 1.61 ± 0.23, *e.g.*, close to 2:1, supporting our hypothesis of a redox reaction between $I^-$ and $Pb^{2+}$. For a more detailed derivation of the ratio, see supporting information section S2.

We observed similar changes in the XPS core level spectra for the blue perovskites, *e.g.*, the formation of $Pb^0$ (Figure 1f), however to a lesser extent compared to the red perovskites. After 20 h of irradiation, a total of 10.2 % $Pb^0$ was observed for the blue perovskites (Figure S3). Additionally, we measured a drastic decrease by ~20–25 % (Figure 1h) in the Cl 2p peak intensity, suggesting chloride as the redox partner for $Pb^{2+}$ since this is roughly twice the amount of $Pb^0$ formed. Finally, the Cs 4d orbital was monitored and no formation of novel peaks could be observed (Figure 1d) + h), suggesting that Cs remains in a relatively stable environment in both cases.

We probed the morphology of both NC samples after X-ray exposure by scanning electron microscopy (SEM). While the red perovskites maintained the cubic, nanocrystalline morphology, the blue perovskites recrystallized to larger agglomerates (see Figure S4). Therefore, we determine the structure before (Figure 2a) and after X-ray exposure by WAXS. (Note that the flux and exposure times during synchrotron-based WAXS are vastly different from those during XPS, and the cross-section is substantially higher for the XPS experiments with their lower photon energies, such that a quantitative comparison is not possible. We believe, however, that qualitative considerations may still be made.) We index the WAXS patterns observed here according to the cubic perovskite phase, although previous scattering experiments have shown a pseudocubic annotation to be more appropriate. [22,23] However, the limited q-range and the broad reflections, which are typical for NCs, prevent such a refinement. From Figure 2 b) one can see that one of the 200 peaks shifted to higher q values, while another remains at the same position. This indicates a contraction of this part of the sample. For further analysis, we integrate the diffraction maps over all angular coordinates to obtain an azimuthally-independent scattering pattern. In Figure 2c), we display the differential intensity, obtained by subtracting the azimuthally integrated diffraction pattern after X-ray exposure from the first pattern. In this Figure, positive differential intensities refer to reflections which weakened during X-ray exposure, while negative differential intensities indicate newly evolved reflections. From the differential pattern, a change of the lattice constant from 6.11 Å to 5.67 Å could be obtained (see Figure 2 c) and Supporting information, S5, for more details).

Upon exposing the red perovskites to X-rays during XPS, we observed a shift of all sample peaks to higher binding energies (Figure 3). We argue that this shift is not primarily caused by electric charging since 1) the Au 4f substrate peaks remain at constant energies throughout the experiment (Figure 3a + b), 2) the perovskite films were rather thin to enable the release of a large number of secondary electrons from the Au substrate underneath to suppress charging, [24,25] and 3) the magnitude of the shift of the sample peaks was different for different elements (Pb 4f: +0.63 eV, Cs 4d: +0.71 eV). The shifts remained constant after illumination for approximately 7 h (Figure 3c + d).

We probed the influence of X-ray irradiation on the work function by measuring UPS before and after 20 h exposure (Figure S5). We find a shift of the high binding cutoff by 0.12 eV to higher energies, indicating a decrease of the work function. At the same time, the energy of the valence state maximum ($1S_h$) referenced to the Fermi level remains constant (see Figure S8). These findings suggest that the energy levels of the nanocrystal are shifted by 0.12 eV toward the vacuum level.

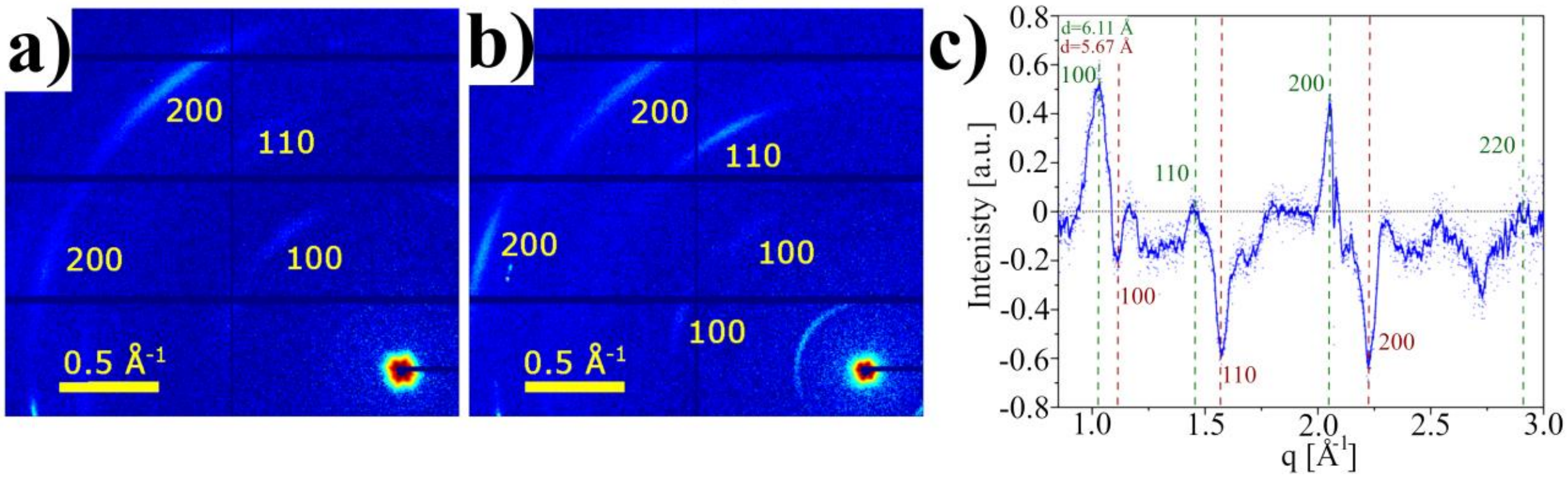

**Figure 2.** a) First WAXS measurement of $CsPbBrI_2$ nanocrystals and b) subsequent measurement after the spot was already exposed to radiation. c) Differential scattering intensity obtained by subtracting the azimuthally integrated diffraction pattern after X-ray exposure from the first pattern. Positive differential intensities refer to reflections which weakened during X-ray exposure, while negative differential intensities indicate newly evolved reflections. Indexing of old (green) and new (brown) peaks according to a cubic perovskite phase.

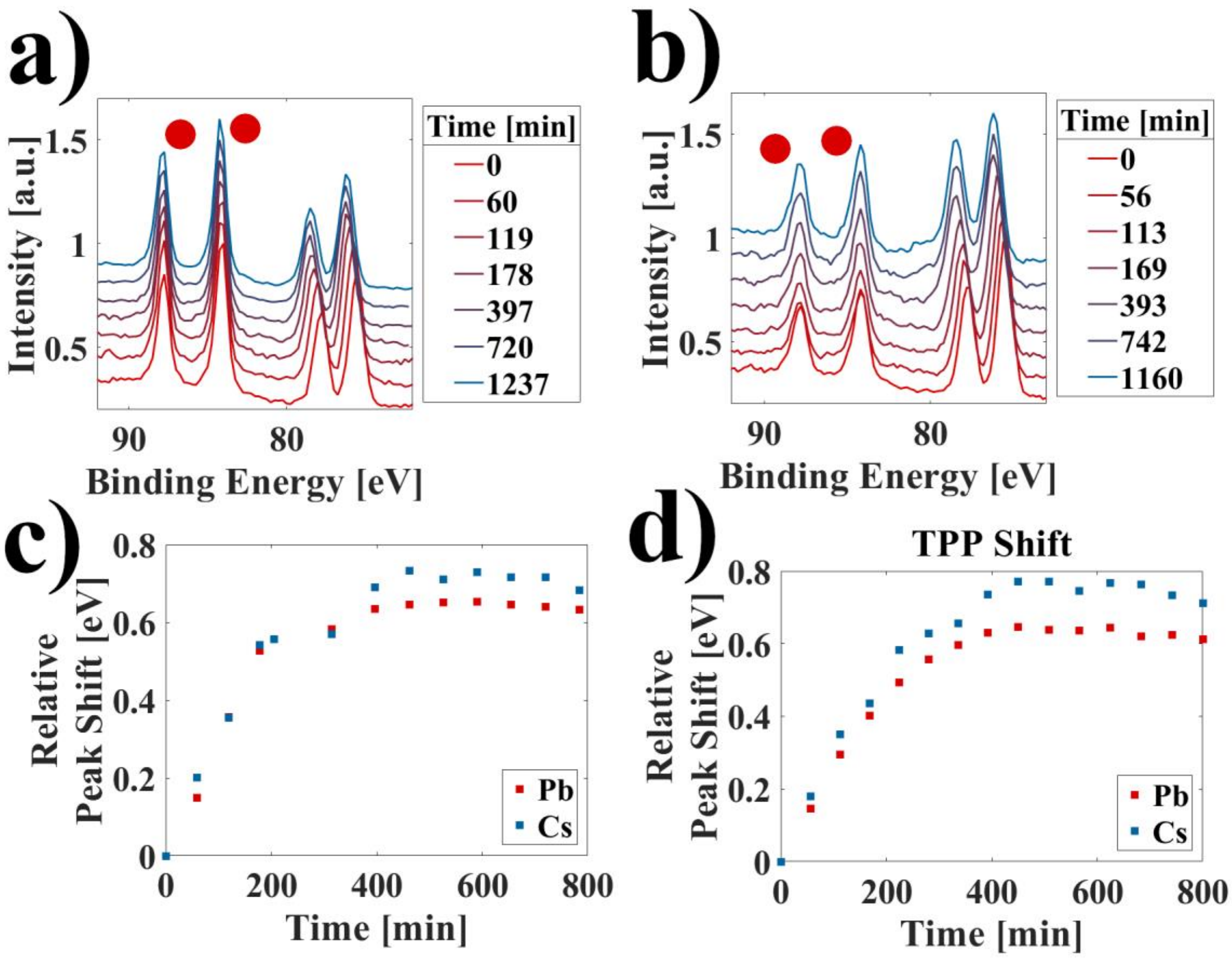

**Figure 3.** Cs 4d (75 eV) and Au 4f (84 eV) XPS spectra for a) native ligands and b) mZnTPP exchanged samples after different X-ray radiation exposure times. The substrate-related Au signals are denoted by red dots and did not change throughout the experiment in both cases. Relative peak position of the Pb 4f and Cs 3d signals compared to the binding energy at the start of the experiment for c) native and d) mZnTPP exchanged samples.

Motivated by the previous finding that ligand exchange with mZnTPP can enhance the stability of red perovskites in light emitting devices, [26] we studied the effect of this ligand on the decomposition of the red perovskites under X-ray illumination. (Note: the loss of structural integrity of the blue perovskites due to chlorine evaporation did not allow for an analogous analysis.) With the mZnTPP ligand shell, we again found the formation of $Pb^0$, but its formation is slower and yields only 8.1 % compared to 13.2 % after 20 h for the native ligand. An additional lead species occurred at higher binding energy for this ligand at ~140.9 eV, indicating a higher oxidation state than $Pb^{2+}$ (Figure 4a + b). The intensity ratio between this new peak and the evolving $Pb^0$ peak in the Pb 4f spectrum is roughly 2:1 for all irradiation times, implying that their formation is correlated (Figure 4e). In contrast to the sample stabilized with the native ligand (*cf.* Figure 1), we did not find a similar shoulder in the iodide signal (Supporting information section S7), suggesting a different decomposition mechanism as a result of the ligand exchange. Before investigating this mechanism in more detail, we verify by SEM the structural integrity of the mZnTPP-stabilized red perovskites after X-ray exposure (Figure 4c) and note that the stoichiometric composition remained roughly constant (Figure 4d). We found the same shift in binding energies for all core-level peaks of the sample as already described for the red perovskites with the native ligand, suggesting that the shift is independent of the ligand shell and related to the NCs itself. UPS measurements before X-ray exposure revealed a shift to lower cutoff binding energies by ~0.5 eV compared to the NCs with native ligand stabilization (Figure S7). After X-ray exposure, the cutoff binding energies increase by 0.1 eV (Figure S5), reproducing the same effect as observed with the native ligand.

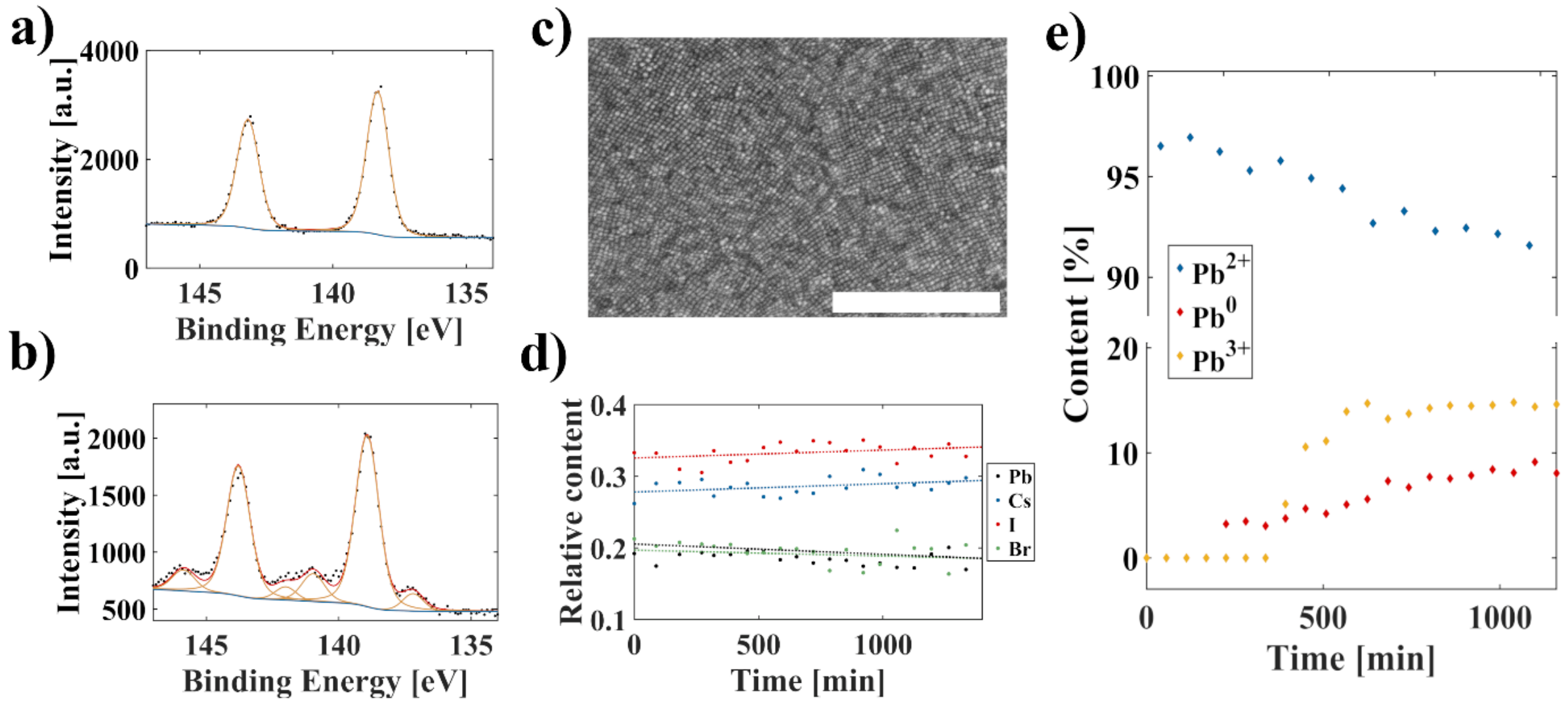

**Figure 4.** XPS of the Pb 4f orbital for $CsPbBrI_2$ with an mZnTPP ligand shell at the a) start and b) end of the experiment. Two new lead species occurred, at lower and higher binding energies. c) SEM micrograph of the mZnTPP exchanged sample after X-ray exposure for approximately 20h. Scale bar corresponds to 500 nm. d) Relative stoichiometric content and linear regression (dotted lines) of the exchanged $CsPbBrI_2$ sample over time. e) Temporal evolution of the observed lead species, the novel formed peaks are attributed to $Pb^0$ (~ 137 eV) and $Pb^{3+}$ (~ 141 eV).

Based on the XPS and WAXS results, we now propose a mechanism for the X-ray radiation-related degradation of the red and blue perovskite NCs with their native ligands. Our findings suggest a redox reaction during which $Pb^0$ and molecular iodine ($CsPbBrI_2$) or chlorine ($CsPbBr_2Cl$) are formed. While iodine resides within the sample after 20 h in ultra-high vacuum, chlorine is removed. This suggests the following decomposition reaction for the red perovskites:

$$\begin{gathered} CsPbBrI_2(xl) \rightarrow CsBr(s) + PbI_2(s) \\ PbI_2(s) \rightarrow Pb^0(s) + I_2(s)\,, \end{gathered} \tag{1}$$

where (xl) refers to the crystalline and (s) to the solid state. Similar mechanisms are postulated for $MAPbI_3$ and $CsPbBrI_2$ thin films, where a comparable Pb peak was found. [6,15–17,20] The first reaction is facilitated by the low enthalpy of formation for $CsPbBrI_2$ as shown by calorimetry. [27] We believe that the second reaction is enabled by the high energy radiation. In view of the high volatility of $I_2$, particular in ultra-high vacuum, we note that the formation of polyiodides, such as $I_3^-$, is possible under these conditions which greatly reduces its volatility. [17,28]

We suggest a similar decay mechanism for the blue perovskites based on the analogous formation of a $Pb^0$ signal (Figure 1f). We interpret the absence of a novel chlorine peak (Figure 1g) and the strong decrease in chlorine content (Figure 1h) as indirect evidence for the oxidation of chloride, since $Cl_2$ is highly volatile and polychlorides are less stable than polyiodides. In this scenario, chlorine would immediately evaporate and remain undetected by XPS. The crystal decomposition expected from such a loss in material is consistent with the greatly altered morphology found by SEM (Figure S4). Our finding that, in the red and blue perovskites alike, bromide is neither oxidized nor removed from the sample, can be rationalized in terms of the oxidation potentials and enthalpies of decomposition for all three halides. Firstly, the oxidation potential of bromide ($E^0$ = -1.087 V) is larger than that of iodide ($E^0$ = -0.5355 V) and triiodide ($E^0$ = -0.536 V), favoring the oxidation of the latter. In contrast, the oxidation potential of chloride ($E^0$ = -1.396 V) is higher than for bromide. However, the enthalpy of decomposition is roughly 0.22 eV larger for $CsPbBr_3$ than for $CsPbCl_3$, which we hold responsible for the observed overall oxidation and removal of chloride. [29] We now discuss the shift of the core-levels in XPS upon X-ray illumination, for which we focus on the red-emitting perovskites since the decomposition of the blue perovskites prevents a similar analysis. A core-level shift in XPS is generally attributed to a variety of origins, such as changes in the chemical environment of specific atoms, an altered electrostatic or Madelung potential, surface effects at the sample substrate interface or charging effects. [7,12–14] Firstly, we rule out any surface effects due to the very prominent gold substrate signal (Figure 3) which remains unaltered throughout the entire experiment. Secondly, charging effects are unlikely due to several reasons: The gold substrate signal is clearly visible, indicating a very thin perovskite film that can be assumed to be grounded. Closely related is the high secondary electron count, originating from the substrate that we correlate to a suppression of sample charging. Lastly, the conductivities of the native as well as the mZnTPP exchanged sample are known. [26] The electric resistance in the porphyrin-containing sample is lower and should result in a smaller shift. However, we find the same shifts independent of the ligand shell and can thus assume that the shift does not originate from charging. Therefore, only two possible explanations remain: changes in either the electrostatic potential or the chemical environment. We argue that a change of the electrostatic potential in the sample is the origin of the peak shifts, which is justified in the following. The electrostatic potential in an ionic solid for an ion $i$ is given as

$$V_i = \frac{e^2}{4\pi\epsilon_0 r_0} \sum_{j\neq i} \frac{z_j}{r_{ij}/r_0} = \frac{e^2}{4\pi\epsilon_0 r_0} M_i \qquad (2)$$

with the elementary charge $e$, vacuum permittivity $\varepsilon_0$, equilibrium lattice constant $r_0$, effective charge $z_j$ of the j-th atom, the distance between the respective atoms $r_{ij}$ and the Madelung constant $M_i$. To quantify the change in $V_i$ for Pb and Cs according to equation (2), we calculate the changes of the radii during the contraction of the unit cell by 0.44 Å as determined from WAXS (Figure 2a) and compare the results to the core-level shifts in XPS. The excellent agreement suggests that the lattice contraction is the reason for the measured core-level shift. We attribute the remaining small discrepancies to the inhomogeneous composition of the mixed halide system, as well as the simultaneously occurring decomposition mechanisms.

**Table 1.** Influence of irradiation by X-rays on the lattice constants and resulting distances for the $CsPbBrI_2$ NCs. The calculated electrostatic potential for Pb and Cs as well as the difference are given, experimental value in brackets.

| | $CsPbBrI_2$ | | | | | |
|---|---|---|---|---|---|---|
| | **$a_0$=6.11Å** | **$a_1$=5.67Å** | **Difference [eV]** | **$a_0$=6.11Å** | **$a_1$=5.67Å** | **Difference [eV]** |
| **Atom** | **Pb** | | | **Cs** | | |
| $d_{CsA}$ [Å] | 5.2395 | 4.8497 | | n.a. | n.a. | |
| $d_{AX}$ [Å] | 3.025 | 2.8 | | 4.2780 | 3.9598 | |
| $d_{PbA}$ [Å] | n.a. | n.a. | | 5.2395 | 4.8497 | |
| $V_{el}$ [eV] | 7.5 | 8.1 | 0.6 (0.630) | 8.9 | 9.6 | 0.7 (0.711) |

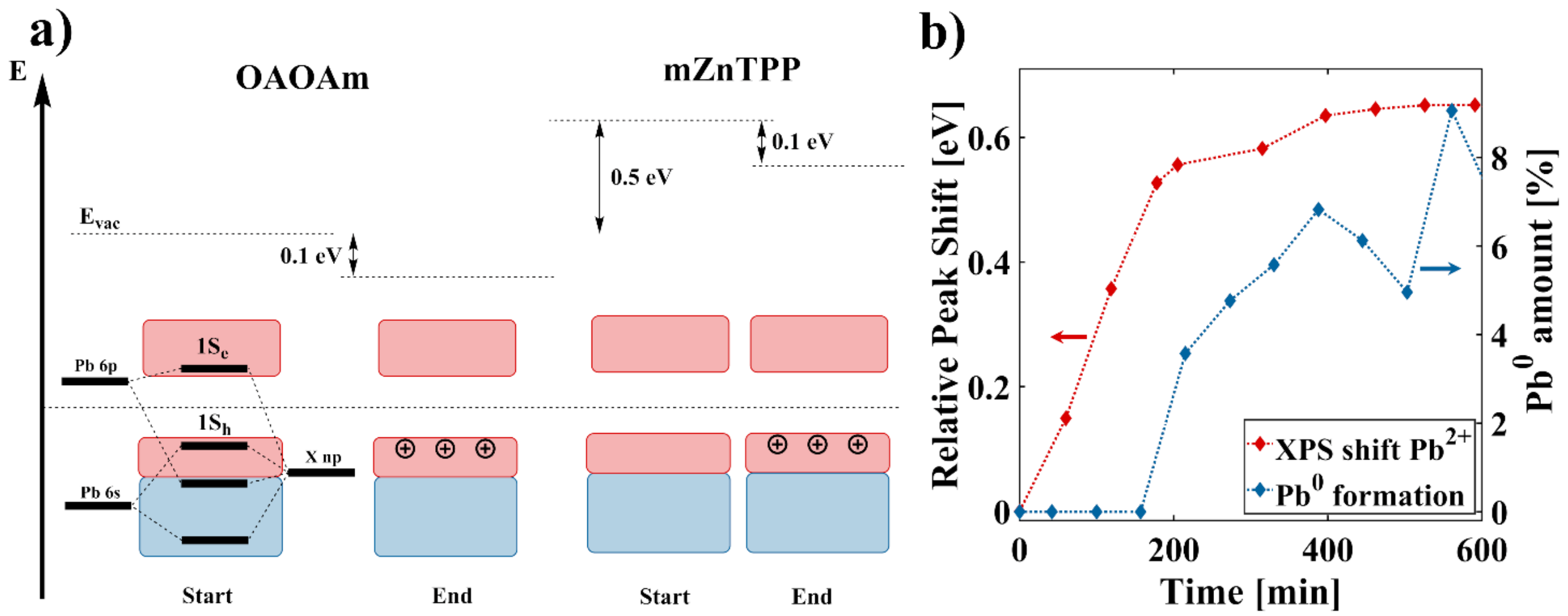

**Figure 5.** a) Energy level scheme obtained from UPS measurements for the native shell (left) and mZnTPP functionalized nanocrystals (right), all energies are referenced against the Fermi level of the instrument. The conduction state minimum ($1S_e$) and valence state maximum ($1S_h$) are indicated, respectively. The character of the states is depicted as red and blue for antibonding and bonding orbitals, respectively. The contraction of the crystal lattice resulted in a shift to higher energies by 0.1 eV for the native ligand as well as for mZnTPP. In addition, mZnTPP functionalization lowered the binding energy by 0.5 eV compared to the native functionalization. b) Temporal evolution of the XPS shift induced by the lattice contraction (red) and the formation of elemental lead (blue).

We now discuss the effect of ligand exchange with mZnTPP on the stability of the red perovskites under X-ray irradiation. Based on our XPS results, which involve the occurrence of two novel lead species (Figure 4b), supposedly $Pb^0$ and $Pb^{3+}$, and a constant $Pb^0$:$Pb^{3+}$ ratio of 1:2 during the experiment (Figure 4e), we suggest the following disproportionation reaction:

$$3\ CsPbBrI_2 \rightarrow Cs_3Pb_2Br_3I_6 + Pb^0 \quad (3)$$

This disproportionation is consistent with our finding that no other iodide species occurred (Figure S9) under these conditions, that is, this degradation pathway does not involve halide oxidation. Despite the low stability of $Pb^{3+}$, this oxidation state has previously been postulated for perovskites. [14,30] In addition, the specialized experimental conditions (continuous X-ray radiation in ultra-high vacuum) may facilitate its detection. Most notably, this new degradation pathway is substantially slower than the degradation with native ligand functionalization (compare Figure S3 with Figure 4e) and proceeds solely via the reaction of lead. To rationalize this surprising effect of the mZnTPP ligand, we note a recent work on the stabilizing effect of ligands with strong (surface) dipoles on CdSe NCs against irreversible reduction during charging. [31] During XPS, the NCs are subject to substantial charging and, while not all considerations for the reduction of Cd in CdSe may be transferable to Pb in $CsPbBrI_2$, we argue that the general rational outlined by du Fossé *et al.* is of central importance also in the present case: mZnTPP invokes a 0.5 eV increase in work function (Figure 5a) and provides better dielectric screening (a higher permittivity) [26] compared to the native ligand. As du Fossé *et al.* have shown, a reduced work function affects primarily the overall crystal and only to a lesser extent a localized state, such as $Pb^0$. [31] This stabilizes the NCs during charging and inhibits the irreversible reduction of lead. Thus, a promising strategy to further enhance the stability of perovskite NCs is the search for ligands that induce even larger work function.

We note that both degradation pathways – with and without mZnTPP – are preceded by the same core level energy shifts of Pb and Cs (Figure 3c+d), which we were able to correlate with a lattice contraction (Figure 2a and Table 1). A likely scenario for such a contraction could be either a phase transition or halide segregation. [16] The latter is a well-known phenomenon in $CsPbI_2Br$, leading to bromine-enriched crystal domains (of smaller lattice constant) with iodine-rich segregations at the boundaries. [17] Figure 5b suggests that this transformation is a prerequisite for the redox reaction of $Pb^{2+}$ to occur according to either equation (1) or (3).

In conclusion, we have shown that the mechanism of photodegradation under X-ray radiation of all-inorganic mixed lead halide perovskite nanocrystals depends on the ligand shell. With the ligands oleic acid/oleylamine, we found a fast decomposition into elemental lead and halogen, similar to previous studies on bulk thin films. After ligand exchange with a metal porphyrin derivative, photodegradation was significantly slower and progressed *via* a disproportionation of $Pb^{2+}$ to $Pb^0$ and $Pb^{3+}$. We hold an increase in work function of the nanocrystal film by 0.5 eV responsible for the altered photodegradation behavior, which was induced by the metal porphyrin derivative. This work highlights the advantageous tunability of the ligand shell of lead halide perovskite nanocrystals as an additional means to improve their photostability and suggests surface ligands that introduce strong dipoles as a general paradigm toward mitigating photodegradation.

AUTHOR INFORMATION

The manuscript was written through contributions of all authors. All authors have given approval to the final version of the manuscript.

ACKNOWLEDGMENT

This work was supported by the DFG under grants SCHE1905/8-1 (project no. 424708673) and SCHE1905/9-1 as well as the Carl Zeiss Stiftung (Forschungsstrukturkonzept "Interdisziplinäres nanoBCP-Lab"). I.A.V. acknowledges the financial support of the Russian Federation represented by the Ministry of Science and Higher Education of the Russian Federation (Agreement No. 075-15-2021-1352). The Authors would like to thank Dmitry Lapkin and Jerome Carnis for help with the WAXS measurements. We acknowledge DESY (Hamburg, Germany), a member of the Helmholtz Association HGF, for the provision of experimental facilities. Parts of this research were carried out at PETRA III synchrotron facility and we would like to thank the beamline staff for assistance in using the Coherence Application P10 beamline.

# Impact of the Ligand Shell on Structural Changes and Decomposition of All-Inorganic Mixed-Halide Perovskite ($CsPbX_3$) Nanocrystals under X-Ray Illumination

## Supplementary Information

*Jan Wahl[$,‡], Philipp Haizmann[$,‡], Christopher Kirsch[$], Rene Frecot[$], Nastasia Mukharamova[#], , Dameli Assalauova[#], Young Yong Kim[#], Ivan Zaluzhnyy[+], Thomas Chassé[$,ß], Ivan A. Vartanyants[#,&], Heiko Peisert[$,]*, Marcus Scheele[$,ß,]**

[$] Institut für physikalische und theoretische Chemie, Universität Tübingen, Auf der Morgenstelle 18, D-72076 Tübingen, Germany

[#] Deutsches Elektronen-Synchrotron DESY, Notkestraße 85, D-22607 Hamburg, Germany

[+] Institut für Angewandte Physik, Universität Tübingen, Auf der Morgenstelle 10, D-72076 Tübingen, Germany

[ß] Center for Light-Matter Interaction, Sensors & Analytics LISA[+], Universität Tübingen, Auf der Morgenstelle 15, D-72076 Tübingen, Germany

[&] National Research Nuclear University MEPhI (Moscow Engineering Physics Institute), Kashirskoe shosse 31, 115409 Moscow, Russia

[‡] These authors contributed equally

*To whom correspondence should be addressed: marcus.scheele@uni-tuebingen.de and heiko.peisert@uni-tuebingen.de

## Materials and methods

### Materials

1-Octadecene (ODE), technical grade, 90%, Sigma Aldrich; Oleic acid (OA), 97%, Acros Organics; Oleylamine (OAm), 80-90%, Acros Organics; Caesium carbonate ($Cs_2CO_3$), 99.99% (trace metal basis), Acros Organics; Lead(II)iodide ($PbI_2$), 99.999% (trace metal basis), Sigma Aldrich; Lead(II)bromide ($PbBr_2$), ≥98%, Sigma Aldrich; Lead(II)chloride ($PbCl_2$), Puratronic™, 99.999% (metal trace), Crystalline, Alfa Aesar; Toluene, HPLC grade, 99.8%; Toluene, 99.8%, extra dry, AcroSeal, Acros Organics; zinc-(5-monocarboxyphenyl-10,15,20-triphenylporphyrin) (mZnTPP), TriPorTech; Tetrachloroethylene (TCE), ≥99%, Acros Organics; Kapton® polyimide membranes (125 μm thickness), DuPont

### $CsPbX_3$ nanocrystal synthesis

The used nanocrystals were synthesized with two different stoichiometries, namely $CsPbBrI_2$ and $CsPbBr_2Cl$, following the published synthesis route by Krieg *et al.* [1] with slight adjustments. For $CsPbBrI_2$ a 20ml glass reaction vial was used which could be heated to the reaction temperature of 160°C in a custom-made aluminum heating block. Generally, the syntheses were carried out with twice the concentration of precursors compared to literature.

### Ligand exchange and thin film preparation

Following the purification, the nanocrystals (NC) were either used as obtained or post-synthetically modified by ligand exchange with zinc-(5-monocarboxyphenyl-10,15,20-triphenylporphyrin) (mZnTPP). The exchange was effectively carried out in solution by adding 0.25 stoichiometric equivalents of mZnTPP to the NC solution. An immediate color change was observed upon addition. The ligand exchange procedure and corresponding analysis is given in more detail in another paper. [2]

The as-synthesized and exchanged NCs were subsequently spin-coated onto custom-made gold substrates under nitrogen atmosphere to prepare thin films. The coating parameters were chosen to be 10 rps for 30 s with a 3 s ramp.

The used substrates were custom made at the LISA$^+$ center Tübingen. A commercially available four-inch silicon wafer with native oxide layer was coated with 10 nm chromium in an evaporation chamber, followed by deposition of a 50 nm gold layer.

## Self-assembly into superlattices

To prepare superlattices for the X-ray scattering measurements, the perovskite samples were prepared as 1-3 mM solutions in toluene ($CsPbBrI_2$) or tetrachloroethylene ($CsPbBr_2Cl$) and drop-casted onto Kapton® substrates under inert atmosphere. The substrates were placed inside a petri dish with an additional reservoir of solvent (1-2 ml) to slow down the evaporation process and covered with a lid. The samples were allowed to dry for 24 h before the lid was removed and an additional drying process for 5–6 h was allowed.

## Scanning electron microscopy (SEM)

SEM was carried out at a HITACHI SU8030 electron microscope, utilizing an acceleration voltage of 30 kV.

## X-ray photoelectron spectroscopy and ultraviolet photoelectron spectroscopy

The thin film samples were analyzed inside an ultrahigh vacuum chamber (base pressure: 2 x $10^{-10}$ mbar) with a XR-50M X-ray source from SPECS utilizing monochromatic Al K$\alpha$ radiation ($h\nu$

= 1486.7 eV). For photoelectron detection a Phoibos 150 DLD hemispherical photoelectron energy analyzer (SPECS) was used. The spectrometer was calibrated to reproduce the binding energy of Au $4f_{7/2}$ (84.0 eV) and Cu $2p_{3/2}$ (932.6 eV) signals, with photoionization cross sections of 0.2511 and 0.3438, respectively. [3] Measurements were performed under fixed analyzer transmission mode, with an energy resolution of 400 meV and 150 meV for XPS and Ultraviolet photoelectron spectroscopy (UPS), respectively. Custom written scripts were used to measure overview and detailed spectra at specified times during the experiment. Peak fitting was done using the Unifit software package. [4] Peaks are expected to show Voigt profile, meaning a convolution of Lorentzian and Gaussian peaks. The background was modeled with an iterative algorithm to apply Shirley background.

UPS was carried out with similar conditions as XPS measurements. A helium ultraviolet source with an energy of 21.22 eV was used in combination with a Phoibos 150 DLD electron analyzer.

## Small- and wide-angle X-ray scattering

The small- and wide-angle X-ray scattering (SAXS and WAXS) were carried out at the Coherence Applications beamline P10 of the PETRA III synchrotron source at the Deutsche Elektronen-Synchrotron (DESY). The X-ray source provided a beam with a wavelength of $\lambda$=0.0898nm or an energy of 13.8 keV, the beam was focused on a spot size of roughly 400x400 $nm^2$ with a focal depth of 0.5 mm. A two-dimensional EIGER X4M (Dectris) detector with 2070x2167 pixels of size 75x75 $\mu m^2$ was used, it was located 412 mm away from the sample plane. The detector was positioned in a way to allow simultaneous measurements of SAXS and WAXS. The exposure time of the samples was 0.5 s. The obtained diffraction patterns were analyzed by Bragg peak assignment and radial profiles could be achieved by averaging over the angular coordinates.

## S1: Low Binding energy survey from XPS

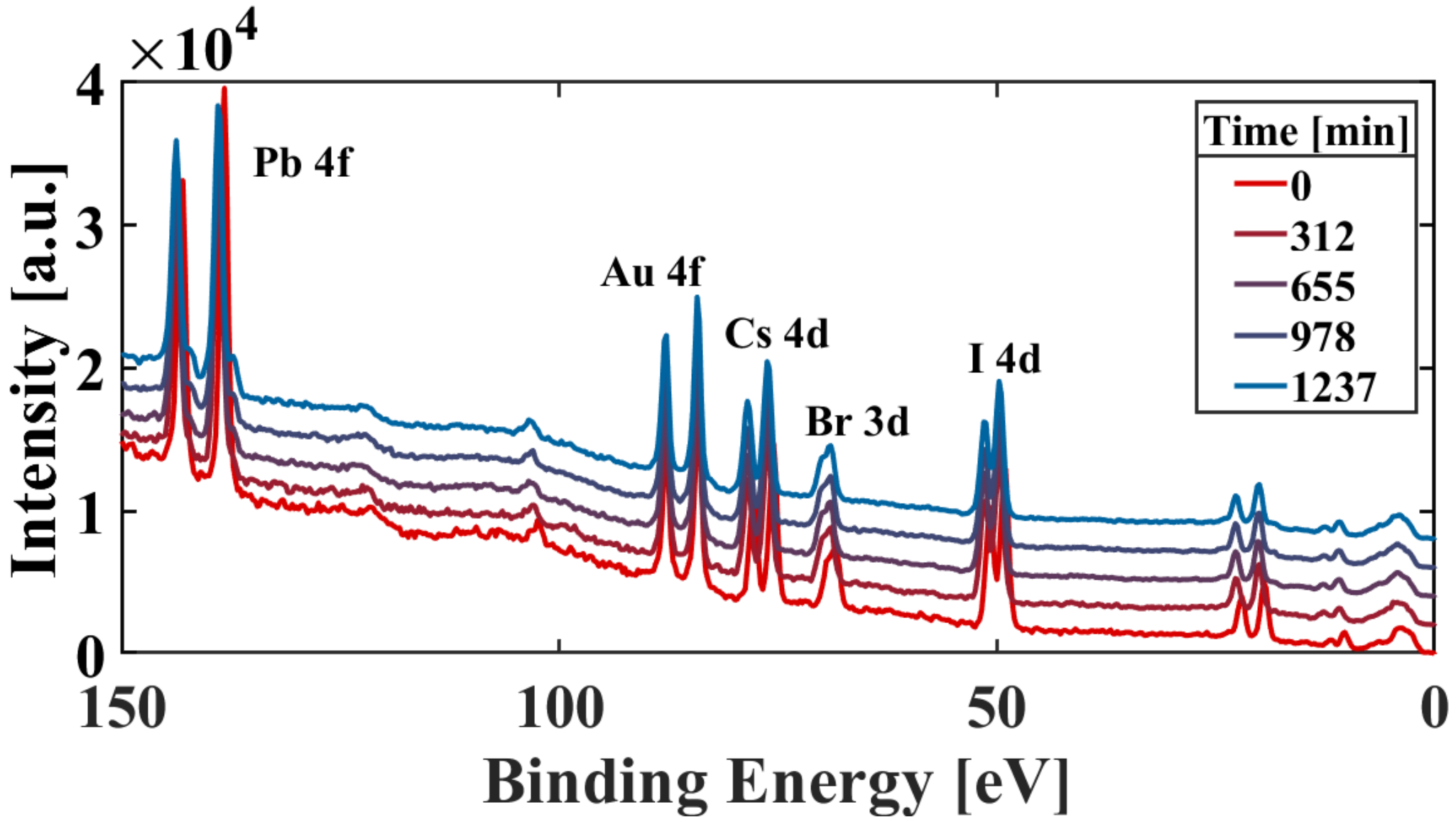

**Figure S1.** Survey spectra at different times during the X-ray illumination. A shift in peak position for all elements associated with the perovskites can be observed. As the perovskites were deposited on a gold substrate, the Au 4f core level peak serves as a reference for the binding energies. Another interesting feature is the occurrence of an additional novel peak in the Pb 4f region, visible at lower binding energies after illumination, indicating the decomposition of the nanocrystals (cf. Figure 1 of the manuscript). Beside of the energetic shift of the Br 3d peak no further change could be observed, *i.e.*, it was not possible to detect an additional species for this element arising from the degradation.

## S2: Calculation of the I:Pb ratio and temporal evolution of novel lead species

As detail XPS spectra are generally not comparable, the ratio of formed elemental lead and iodine was calculated as follows. The percentages were taken from the ratio of the areas fitted to the detail spectra as shown in Figure S2. Subsequently, the areas from the survey spectra were fitted as described in the methods section. The as obtained areas were then multiplied with the percentages from the detail spectra, resulting in areas that are quantitatively comparable.

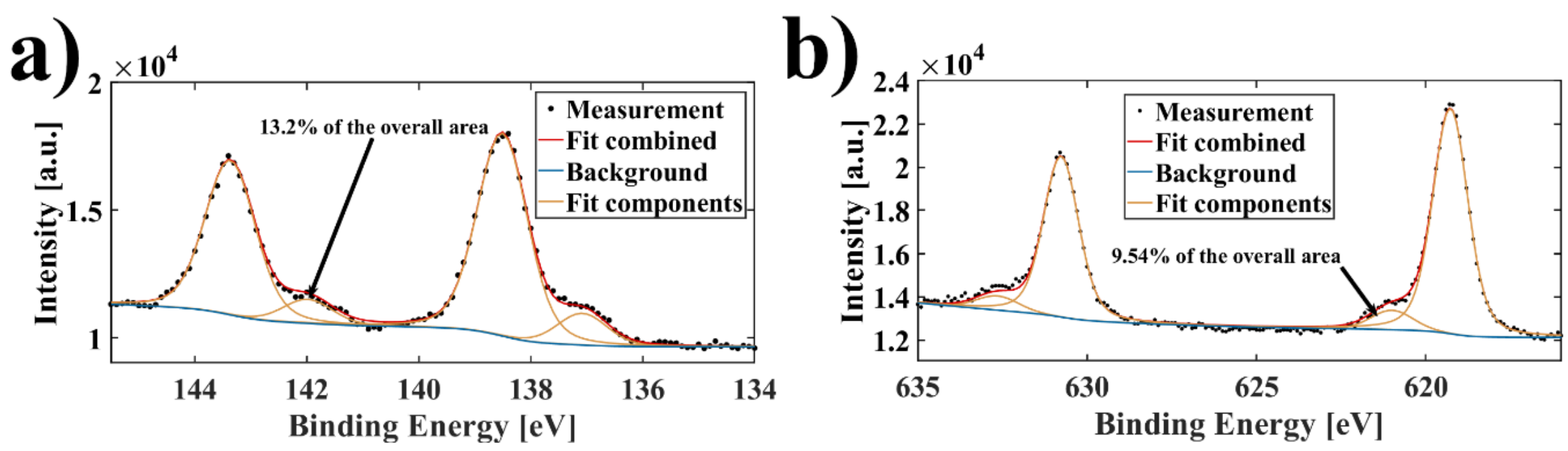

**Figure S2.** Percentages of the novel formed species after illuminating the native sample for ~20h for a) lead and b) iodide. The amount of newly formed species is given as a percentage of the overall peak area in both cases.

Generally, an error of 10% per fitted peak area is assumed. [5] Additionally, a Gaussian error propagation was carried out since multiple calculations were done with the measured values. The precise error calculation is given by:

$$\Delta\left(\frac{I_2}{Pb^0}\right) = \Delta R = \sqrt{\left(\frac{\partial R}{\partial A_{I_2}} * \Delta A_{I_2}\right)^2 + \left(\frac{\partial R}{\partial A_{Pb^0}} * \Delta A_{Pb^0}\right)^2} \qquad \text{(S1)}$$

With the ratio of formed iodine to elemental lead $R$, the Area of iodine $A_{I2}$ and elemental lead $A_{Pb0}$ as well as the corresponding uncertainties denoted by $\Delta$. The error calculation resulted in an error of the ratio of 0.23.

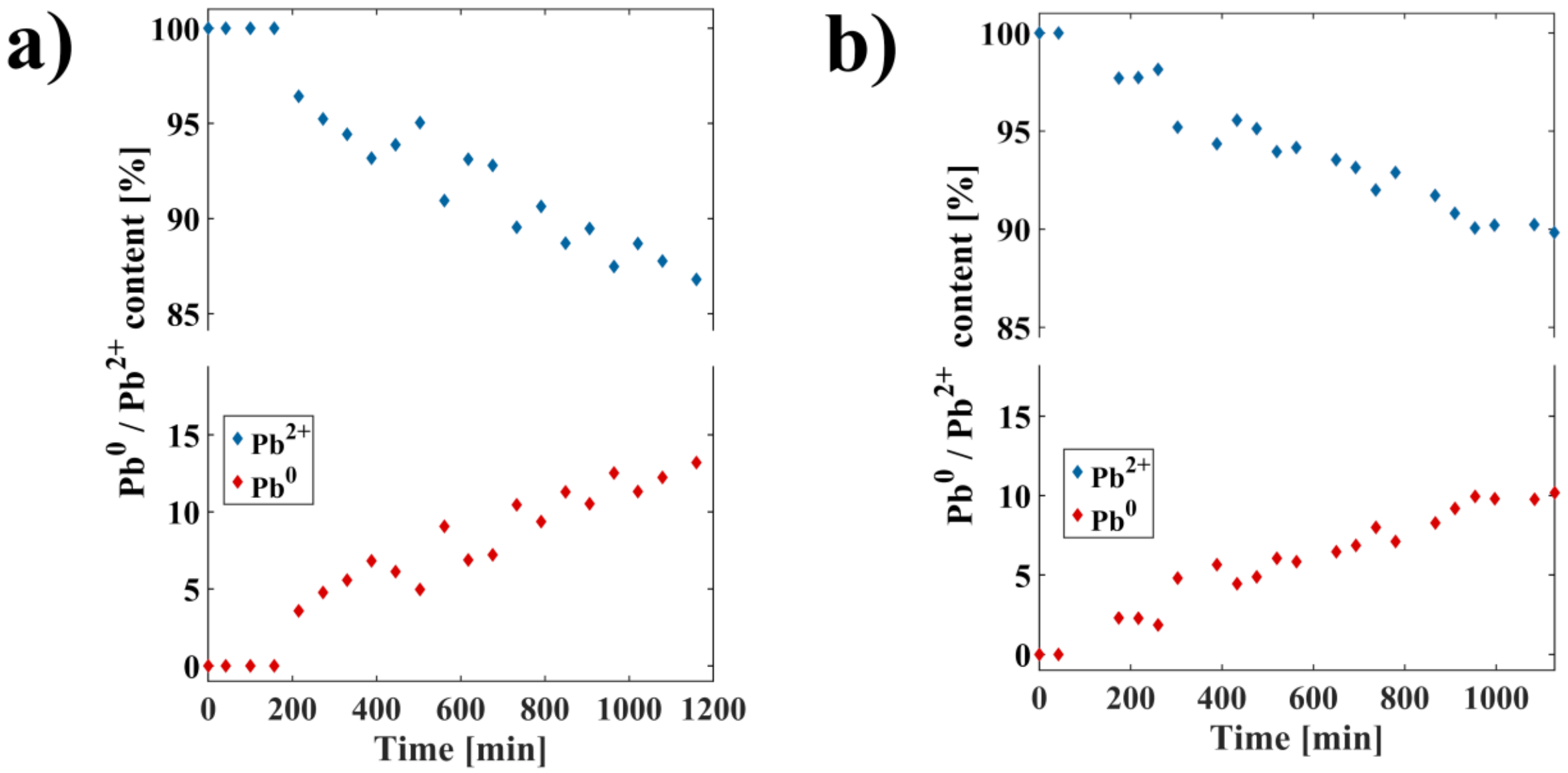

**Figure S3.** Temporal evolution of elemental lead ($Pb^0$) for a) $CsPbBrI_2$ and b) $CsPbBr_2Cl$ obtained by fitting the Pb 4f XPS spectra in Figure 1 of the manuscript.

## S3: Structural integrity probed by SEM for $CsPbBrI_2$ and $CsPbBr_2Cl$

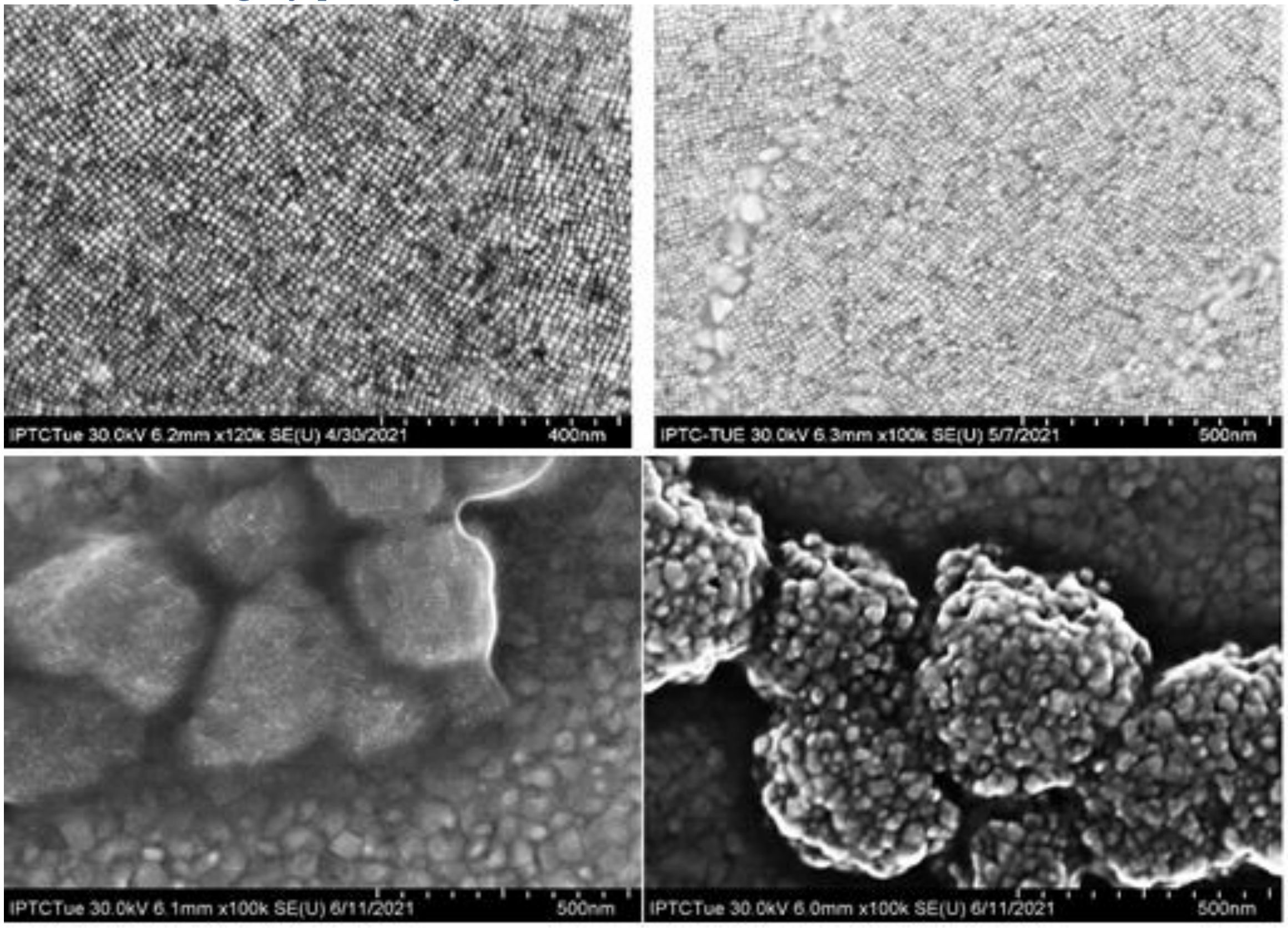

**Figure S4.** SEM of CsPbBrI2 (top) and CsPbBr2Cl (bottom) before (left) and after (right) X-ray exposure.

## S4: UPS of native and exchanged $CsPbBrI_2$ before and after X-ray illumination

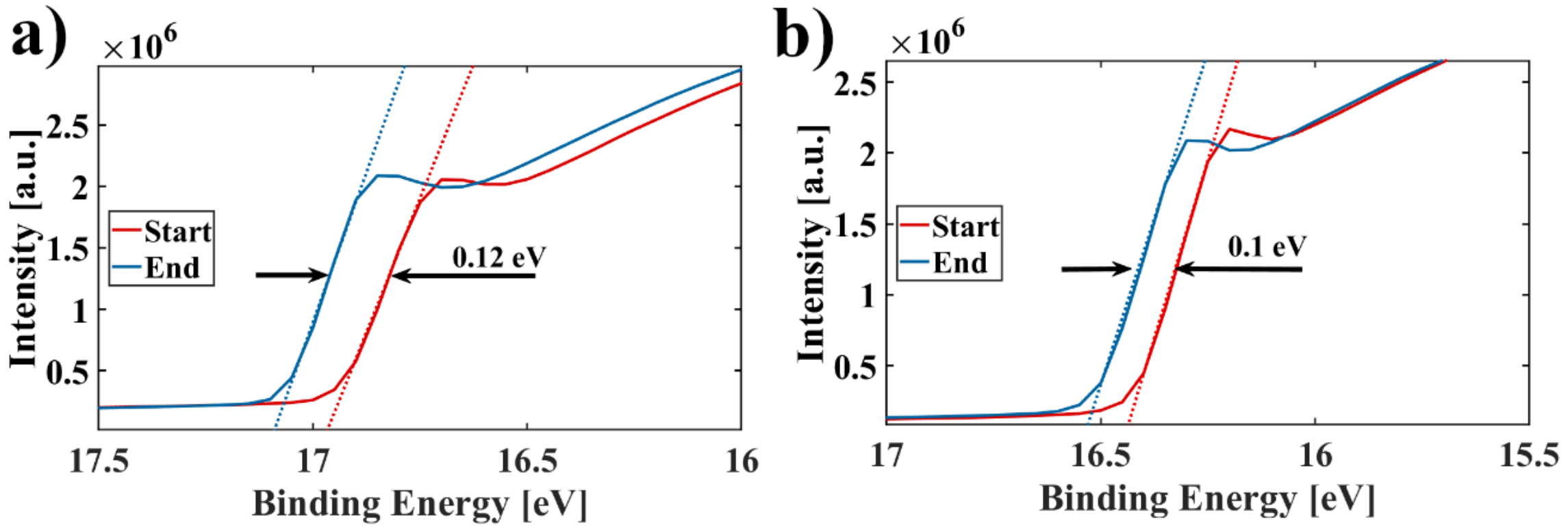

**Figure S5.** UPS cut-off energies of a) native and b) exchanged $CsPbBrI_2$. The shift in the cut-off energy of ~0.1 eV in both cases is found from a linear extrapolation, the shift indicates a reduction of the work function by the same amount. The fits are indicated as dotted lines.

## S5: Further example of the lattice contraction

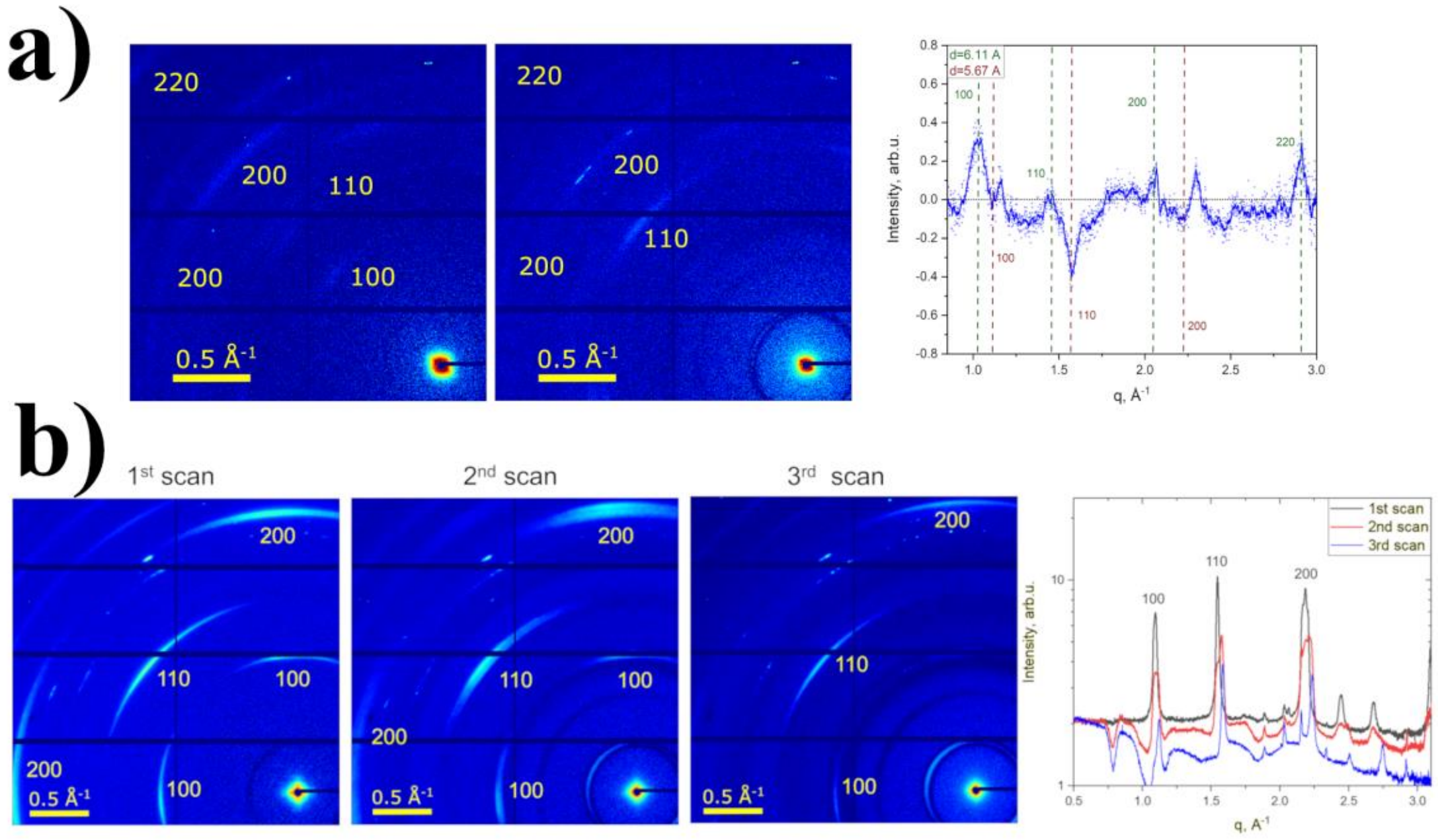

**Figure S6**. More examples of the isotropic lattice contraction for a) $CsPbBrI_2$ and b) $CsPbBr_2Cl$

## S6: UPS overview spectra and onset

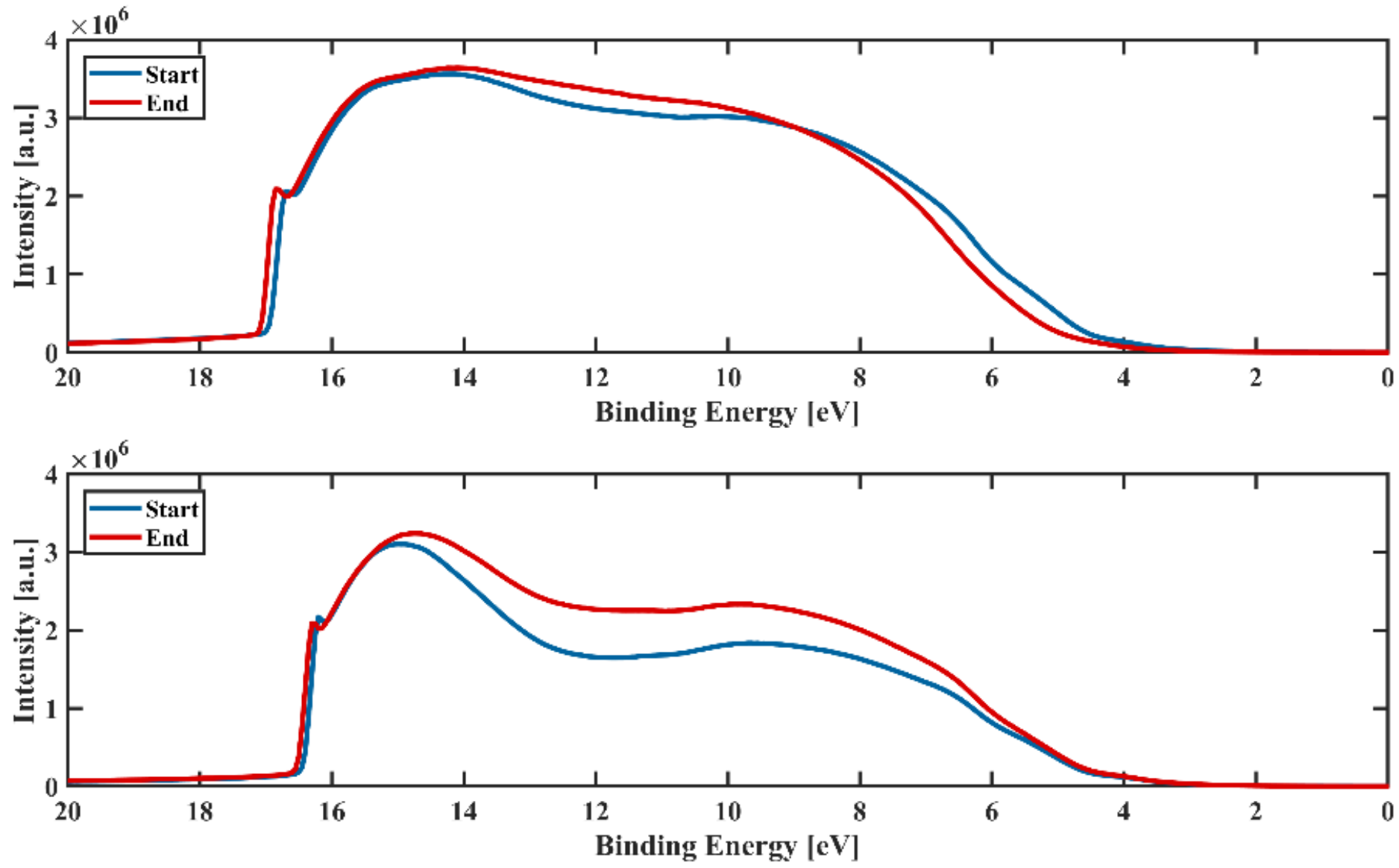

**Figure S7**. UPS overview spectra of the native $CsPbBrI_2$ sample (top) and the mZnTPP exchanges sample (bottom). The cutoff energies are shifted by ~0.5 eV.

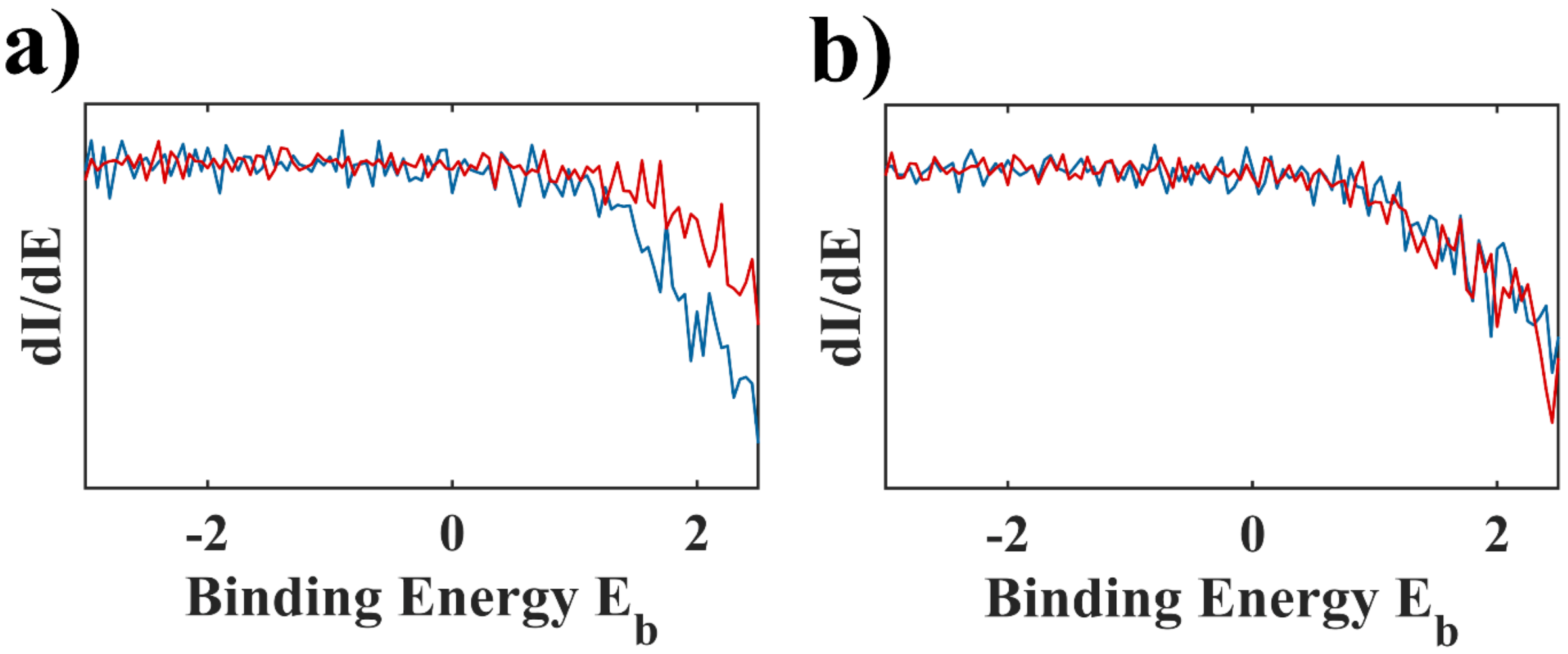

**Figure S8.** The onset remained constant at ~1 eV, shown by the derivative of the UPS spectrum for a) native and b) mZnTPP functionalized $CsPbBrI_2$. The constant onset refers to the Fermi level being located in the middle of the band gap before and after the illumination. Therefore, a shift of the energy states in their entirety is at hand.

## S7: Iodide preservation for CsPbBrI2 functionalized with mZnTPP

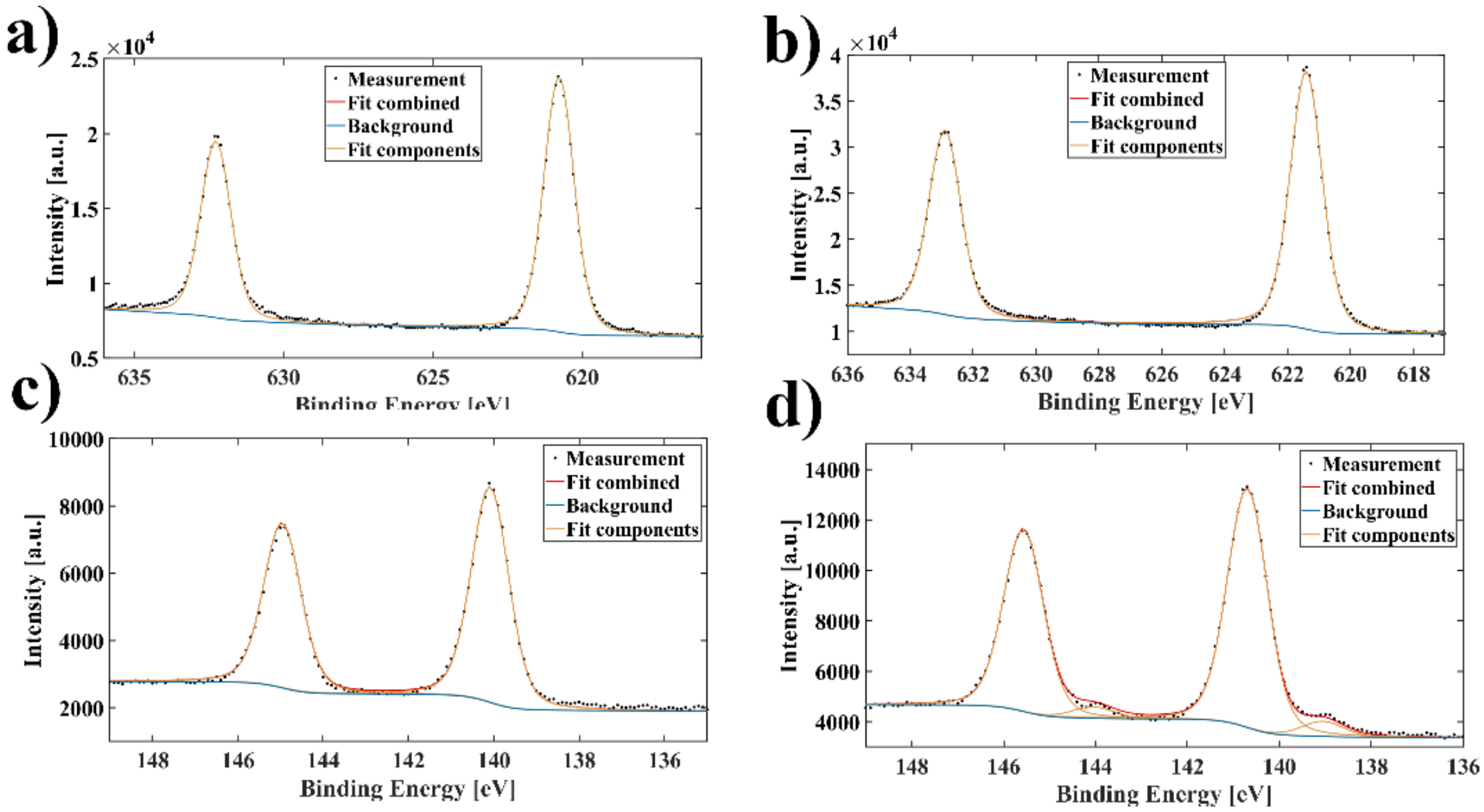

**Figure S9.** I 3d (a + b) and Pb 4f (c + d) XPS detail spectra for the mZnTPP exchanged sample. The decomposition of the NCs started as indicated by the formation of elemental lead. However, there are no novel peaks for the iodide signal.